\documentstyle[12pt]{article}
\newlength{\pubnumber} 

\catcode`\@=11
\@addtoreset{equation}{section}

\def\section{\@startsection{section}{1}{\z@}{3.5ex plus 1ex minus .2ex}
 {2.3ex plus .2ex}{\large\bf}}
\def\subsection{\@startsection{subsection}{2}{\z@}{2.3ex plus .2ex}
 {2.3ex plus .2ex}{\bf}}

\begin{document}

\begin{titlepage}
\samepage{
\setcounter{page}{1}
\rightline{OUTP--00--36P}
\rightline{\tt hep-th/0109162}
\rightline{July 2001}
\vfill
\begin{center}
 {\Large \bf String Inspired ${Z'}$ Model With  \\ 
  Stable Proton and Light Neutrino Masses}
\vfill
\vspace{.25in}
 {\large Alon E. Faraggi\footnote{faraggi@thphys.ox.ac.uk}$^{1,2}$
and Marc Thormeier\footnote{thor@thphys.ox.ac.uk}$^1$\\}
\vspace{.25in}
 {\it  $^1$Theoretical Physics Department,\\
              University of Oxford, Oxford, OX1 3NP, United Kingdom\\}
\vspace{.25in}
{\it $^2$Theory Division, CERN, CH--1211 Geneva, Switzerland} 

\vspace{.05in}
\end{center}
\vfill
\begin{abstract}
  {\rm
Grand unification and its incarnation in the form of heterotic--string
unification, are the only extensions of the Standard Model that are
rooted in the structure of the Standard Model itself. In this context
it was proposed
that the exclusiveness of proton stability and suppression of neutrino masses
necessitates the existence of an
additional $U(1)_{Z^\prime}$ symmetry, which is of non--GUT origin
and remains unbroken down to intermediate or low energies. 
Realistic string models frequently give rise to non--GUT
$U(1)$ symmetries, which arise from the flavor symmetries in the models.
In this paper we demonstrate in a string--inspired toy model
that such a stringy $Z^\prime$ can indeed guarantee proton
longevity and viable phenomenology in the neutrino sector
as well as in the quark and charged lepton sectors. 
}
\end{abstract}
\vfill
\smallskip}
\end{titlepage}

\setcounter{footnote}{0}

\def\beq{\begin{equation}}
\def\eeq{\end{equation}}
\def\beqn{\begin{eqnarray}}
\def\eeqn{\end{eqnarray}}
\def\Tr{{\rm Tr}\,}
\def\KM{{Ka\v{c}-Moody}}

\def\ie{{\it i.e.}}
\def\etc{{\it etc}}
\def\eg{{\it e.g.}}
\def\half{{\textstyle{1\over 2}}}
\def\third{{\textstyle {1\over3}}}
\def\quarter{{\textstyle {1\over4}}}
\def\m{{\tt -}}
\def\p{{\tt +}}

\def\rep#1{{\bf {#1}}}
\def\slash#1{#1\hskip-6pt/\hskip6pt}
\def\slk{\slash{k}}
\def\GeV{\,{\rm GeV}}
\def\TeV{\,{\rm TeV}}
\def\y{\,{\rm y}}
\def\SM{Standard-Model }
\def\SUSY{supersymmetry }
\def\SSM{supersymmetric standard model}
\def\vev#1{\left\langle #1\right\rangle}
\def\l{\langle}
\def\r{\rangle}

\def\Htw{{\tilde H}}
\def\chibar{{\overline{\chi}}}
\def\qbar{{\overline{q}}}
\def\ibar{{\overline{\imath}}}
\def\jbar{{\overline{\jmath}}}
\def\Hbar{{\overline{H}}}
\def\Qbar{{\overline{Q}}}
\def\abar{{\overline{a}}}
\def\alphabar{{\overline{\alpha}}}
\def\betabar{{\overline{\beta}}}
\def\tautwo{{ \tau_2 }}
\def\calF{{\cal F}}
\def\calP{{\cal P}}
\def\calN{{\cal N}}
\def\smallmatrix#1#2#3#4{{ {{#1}~{#2}\choose{#3}~{#4}} }}
\def\bone{{\bf 1}}
\def\V{{\bf V}}
\def\b{{\bf b}}
\def\N{{\bf N}}
\def\bQ{{\bf Q}}
\def\t#1#2{{ \Theta\left\lbrack \matrix{ {#1}\cr {#2}\cr }\right\rbrack }}
\def\C#1#2{{ C\left\lbrack \matrix{ {#1}\cr {#2}\cr }\right\rbrack }}
\def\tp#1#2{{ \Theta'\left\lbrack \matrix{ {#1}\cr {#2}\cr }\right\rbrack }}
\def\tpp#1#2{{ \Theta''\left\lbrack \matrix{ {#1}\cr {#2}\cr }\right\rbrack }}


\def\inbar{\,\vrule height1.5ex width.4pt depth0pt}

\def\IC{\relax\hbox{$\inbar\kern-.3em{\rm C}$}}
\def\IQ{\relax\hbox{$\inbar\kern-.3em{\rm Q}$}}
\def\IR{\relax{\rm I\kern-.18em R}}
 \font\cmss=cmss10 \font\cmsss=cmss10 at 7pt
\def\IZ{\relax\ifmmode\mathchoice
 {\hbox{\cmss Z\kern-.4em Z}}{\hbox{\cmss Z\kern-.4em Z}}
 {\lower.9pt\hbox{\cmsss Z\kern-.4em Z}}
 {\lower1.2pt\hbox{\cmsss Z\kern-.4em Z}}\else{\cmss Z\kern-.4em Z}\fi}

\def\AEF{A.E. Faraggi}
\def\KRD{K.R. Dienes}
\def\JMR{J. March-Russell}
\def\NPB#1#2#3{{\it Nucl.\ Phys.}\/ {\bf B#1} (#2) #3}
\def\PLB#1#2#3{{\it Phys.\ Lett.}\/ {\bf B#1} (#2) #3}
\def\PRD#1#2#3{{\it Phys.\ Rev.}\/ {\bf D#1} (#2) #3}
\def\PRL#1#2#3{{\it Phys.\ Rev.\ Lett.}\/ {\bf #1} (#2) #3}
\def\PRT#1#2#3{{\it Phys.\ Rep.}\/ {\bf#1} (#2) #3}
\def\MODA#1#2#3{{\it Mod.\ Phys.\ Lett.}\/ {\bf A#1} (#2) #3}
\def\IJMP#1#2#3{{\it Int.\ J.\ Mod.\ Phys.}\/ {\bf A#1} (#2) #3}
\def\nuvc#1#2#3{{\it Nuovo Cimento}\/ {\bf #1A} (#2) #3}
\def\etal{{\it et al,\/}\ }

\hyphenation{su-per-sym-met-ric non-su-per-sym-met-ric}
\hyphenation{space-time-super-sym-met-ric}
\hyphenation{mod-u-lar mod-u-lar--in-var-i-ant}


\setcounter{footnote}{0}
\section{Introduction}

Following the astounding experimental achievements of the previous decade,
the Standard Model of particle physics is firmly established
as the correct effective description of nature in the accessible
energy range. The experimentally observed multiplet structure of
the Standard Model, as well as other qualitative indications,
strongly favor the embedding of the Standard Model in larger
unifying groups. This embedding can be realized either at the 
level of an effective field theory below the Planck scale, or 
directly in a heterotic--string realization, formulated as a model
of perturbative quantum gravity at the Planck scale. 

A general consequence of the Standard Model
multiplet unification is that the unification scale is separated 
{}from the electroweak scale by fifteen orders of magnitude.
This arises from the fact that the multiplet unification of the
Standard model spectrum results in proton decay mediating operators.
Compatibility of this hypothesis with the measured gauge parameters
at the electroweak scale and the observed suppression of the 
left--handed neutrino masses then provide further strong support for
the validity of the grand unification scenario. 
The unification hypothesis substantially reduces the number of free
parameters of the Standard Model and is hence also well motivated 
{}from the perspective of reducing some of the arbitrariness found
in the Standard Model. String theory then provides a consistent
framework for quantum gravity unification, in the context of
which all the Standard Model parameters should, in principle,
be calculable from a minimal set of fundamental parameters. 

An important augmentation to the unification program is that of
space--time supersymmetry. While not yet substantiated experimentally,
supersymmetry is supported by several 
observations, including the consistency of a large top quark mass
with the supersymmetric radiative electroweak symmetry breaking.
However, proton longevity is more problematic in supersymmetric
extensions of the Standard Model \cite{nilles}, which admit dimension
four and five baryon and lepton number violating operators \cite{WSY}.
In the MSSM one imposes a global symmetry, $R-$parity ($R_p$),
which forbids the dangerous dimension four operators, while the
difficulty with the dimension five operators can only be circumvented if
one further assumes that the relevant Yukawa couplings are sufficiently
suppressed. The proton lifetime problem becomes especially acute in the
context of heterotic--string unification, in which all operators that
are compatible with the local gauge symmetries are expected to arise from
nonrenormalizable terms. Indeed this issue has been examined in the past by
a number of authors \cite{psinsm,ps94,cus,pati,zprime}. The avenues
explored range from the existence of matter parity at special
points in the moduli space of particular models to the emergence
of non--Abelian custodial symmetries in specific compactifications \cite{cus}.

The most realistic heterotic--string models constructed to date are 
the models constructed in the free fermionic formulation \cite{fff}.
This has given rise to a large set of semi--realistic
models \cite{fsu5,fny,pssm,eu,nahe,cus,cfn,cfs},
which differ in their detailed phenomenological 
characteristics and share an underlying $Z_2\times Z_2$
orbifold structure \cite{foc}.
Past investigations examined several possibilities to
explain the proton longevity. For example, ref. \cite{lepzp}
stipulated the possibility that the $U(1)_{Z^\prime}$, which 
is embedded in $SO(10)$, remains unbroken down to the TeV
scale and suppresses the problematic dimension 4 operators.
In ref. \cite{cus}
it was shown that the free fermionic string models occasionally
give rise to non--Abelian custodial symmetries, which forbid 
proton decay mediating operators to all orders of non--renormalizable
terms. These proposals, however, fall short of providing 
a satisfactory solution, as they
are, in general, exclusive to the generation of light neutrino
masses through a see--saw mechanism. For example, the absence
of the $SO(10)$ 126 representation in string models
\cite{dienes} necessitates
that the $SO(10)$ $U(1)_{Z^\prime}$ be broken at a high scale,
rather than at a low scale. Similarly, to date, the existence 
of the custodial non--Abelian symmetries seems to be exclusive
to the generation of a see--saw mass matrix. 

Refs. \cite{pati,zprime} proposed that heterotic--string
unification necessitates the existence of an additional
$U(1)$ symmetry, beyond the Standard Model,
which remains unbroken down to low or intermediate
energy. Invariance under the extra $U(1)$ forbids the proton decay 
mediating operators, which can be generated only
after $U(1)_{Z^\prime}$ breaking. The magnitude 
of the proton decay mediating operators is therefore
proportional to the $U(1)_{Z^\prime}$ breaking scale,
$\Lambda_{Z^\prime}$, which is in turn constrained
by the proton lifetime limit, and other phenomenological constraints.
On the other hand the additional $U(1)$ should not
forbid quark, lepton and seesaw mass terms.
It was argued \cite{pati,zprime} that the required $U(1)$
symmetry is not of the type that arises in $SO(10)$ or $E_6$ GUTs. 
Rather, it should arise from the $U(1)$ symmetries in the string models 
that are external to the GUT symmetries.

By studying the spectrum and symmetries of the string model
of ref. \cite{eu} Pati showed \cite{pati} that $U(1)$ symmetries
with the required properties do indeed exist in the string models, 
and arise from combinations of the flavor symmetries. 
In ref. \cite{zprime} it was shown that the required
symmetries can in fact remain unbroken by the choices
of supersymmetric flat directions in the string vacuum. 
In the present paper we continue to explore these ideas.
The basic approach that we pursue is to develop field theory 
models that are inspired from the string models. By this 
we aim to confront the string features with the experimental data.
Our main goal in the present paper is to examine whether 
the string symmetries that are used to guarantee the proton
longevity can indeed at the same time allow for adequate suppression 
of the neutrino masses. That this is a nontrivial requirement
can be seen from the fact that if we insisted on gauged $B-L$ 
as the proton protector, it would imply a neutrino in the MeV range
\cite{tauneut}. 

We note that the possibility of utilizing $U(1)$ gauge symmetry
to insure proton longevity has also been discussed in the context
of purely field theory models \cite{ftmoez}. 
The novelty here is the consideration of  $U(1)$ symmetries that 
arise in the context the realistic free fermionic string models.
{}From the string models we extract the charges of the Standard Model
fields under the various $U(1)$ symmetries.
We examine which combinations of the $U(1)$ symmetries forbid the
dangerous baryon and lepton number violating operators while still
allowing for the see--saw mass terms. We then assume that all but two
of the additional $U(1)$s are broken in the string vacuum
by the choices of supersymmetric flat directions. The first
one being the $U(1)_{Z^\prime}$ which is embedded in $SO(10)$,
and the second being the additional unbroken $U(1)$ that
arises from the flavor symmetries of the string model.
We proceed to construct the see--saw mass matrix by
postulating additional Higgs spectrum and comment how it may
arise from the string models. We then demonstrate that
such additional stringy $U(1)$ symmetries can indeed guarantee proton
longevity and viable phenomenology in the neutrino sector,
as well as in the quark and charged lepton sectors. 

\section{Gauge symmetries in free fermionic models}\label{stringmodels}

In this section we discuss the general structure of the realistic
free fermionic models, and of the additional $U(1)$ symmetries
that arise in these models. The
free fermionic heterotic--string formulation yields a large number 
of three generation models, which possess an underlying $Z_2\times Z_2$
orbifold structure \cite{foc}, and differ in their detailed phenomenological 
characteristics. We emphasize
the features of the models that are common to this 
large class of realistic models. 

The free fermionic models are constructed by specifying a set
of boundary conditions basis vectors and the one--loop
GSO projection coefficients \cite{fff}.
The physical massless states are obtained by acting on the vacuum with 
bosonic and fermionic operators and by
applying the generalized GSO projections. The $U(1)$
charges, $Q(f)$, with respect to the unbroken Cartan generators of the four 
dimensional gauge group, which are in one 
to one correspondence with the $U(1)$
currents ${f^*}f$ for each complex fermion $f$, are given by:
\beqn
{Q(f) = {1\over 2}\alpha(f) + F(f)},
\label{u1charges}
\eeqn
where $\alpha(f)$ is the boundary condition of the world--sheet
fermion $f$ in the sector $\alpha$, and 
$F_\alpha(f)$ is a fermion number operator. 

The four dimensional gauge group in the three generation
free fermionic models arises as follows. The models can 
in general be regarded as constructed in two stages.
The first stage consists of the NAHE set, 
$\{{\bf1},S,b_1,b_2,b_3\}$ \cite{nahe}. 
The gauge group after imposing the GSO projections induced
by the NAHE set basis vectors is $SO(10)\times SO(6)^3\times E_8$
with $N=1$ supersymmetry. The space--time vector bosons that generate
the gauge group arise from the Neveu--Schwarz sector and
{}from the sector ${\bf1}+b_1+b_2+b_3$. The Neveu--Schwarz sector
produces the generators of $SO(10)\times SO(6)^3\times SO(16)$.
The sector $\zeta\equiv{\bf1}+b_1+b_2+b_3$ produces the spinorial  
${\bf128}$
of $SO(16)$ and completes the hidden gauge group to $E_8$.
At the level of the NAHE set the sectors $b_1$, $b_2$ and $b_3$
produce 48 multiplets, 16 from each, in the $16$ 
representation of $SO(10)$. The states from the sectors $b_j$
 are singlets of the hidden $E_8$ gauge group and transform 
under the horizontal $SO(6)_j$ $(j=1,2,3)$ symmetries. This structure
is common to all the realistic free fermionic models. At this stage 
we anticipate that the $SO(10)$ group gives rise to the Standard Model 
group factors, whereas the $SO(6)^3$ groups may produce additional 
symmetries that can play a role in safeguarding the proton lifetime. 

The second stage of the free fermionic
basis construction consists of adding to the 
NAHE set three (or four) additional boundary condition basis vectors. 
These additional basis vectors reduce the number of generations
to three chiral generations, one from each of the sectors $b_1$,
$b_2$ and $b_3$, and simultaneously break the four dimensional
gauge group. The $SO(10)$ symmetry is broken to one of its subgroups
$SU(5)\times U(1)$ \cite{fsu5}, $SO(6)\times SO(4)$ \cite{pssm},
$SU(3)\times SU(2)^2\times U(1)$ \cite{cfs}
or $SU(3)\times SU(2)\times U(1)^2$ \cite{fny,eu,cfn}.
Similarly, the hidden $E_8$ symmetry is broken to one of its
subgroups by the basis vectors which extend the NAHE set.
This hidden $E_8$ subgroup may, or may not, contain $U(1)$ factors
which are not enhanced to a non--Abelian symmetry. As
the Standard Model states are not charged with respect to these
$U(1)$ symmetries, they cannot play a role in suppressing the 
proton decay mediating operators, and are therefore not discussed
further here. On the other
hand, the flavor $SO(6)^3$ symmetries in the NAHE--based models
are always broken to flavor $U(1)$ symmetries, as the breaking
of these symmetries is correlated with the number of chiral
generations. Three such $U(1)_j$ symmetries are always obtained
in the NAHE based free fermionic models, from the subgroup
of the observable $E_8$, which is orthogonal to $SO(10)$.
These are produced by the world--sheet currents ${\bar\eta}{\bar\eta}^*$
($j=1,2,3$), which are part of the Cartan sub--algebra of the
observable $E_8$. Additional unbroken $U(1)$ symmetries, denoted
typically by $U(1)_j$ ($j=4,5,...$), arise by pairing two real
fermions from the sets $\{{\bar y}^{3,\cdots,6}\}$,
$\{{\bar y}^{1,2},{\bar\omega}^{5,6}\}$ and
$\{{\bar\omega}^{1,\cdots,4}\}$. The final observable gauge
group depends on the number of such pairings. 

Subsequent to constructing the basis vectors and extracting the massless
spectrum the analysis of the free fermionic models proceeds by
calculating the superpotential. 
The cubic and higher-order terms in the superpotential 
are obtained by evaluating the correlators
\beq
A_N\sim \langle V_1^fV_2^fV_3^b\cdots V_N^b\rangle,
\label{supterms}
\eeq
where $V_i^f$ $(V_i^b)$ are the fermionic (scalar) components
of the vertex operators, using the rules given in~\cite{kln}.
Typically, 
one of the $U(1)$ factors in the free-fermion models is anomalous,
and generates a Fayet--Ilioupolos term which breaks supersymmetry
at the Planck scale. The anomalous $U(1)$ is broken, and supersymmetry
is restored, by a non--trivial VEV for some scalar
field that is charged under the anomalous $U(1)$.
Since this field is in general also charged with respect
to the other anomaly-free $U(1)$ factors, some non-trivial
set of other fields must also get non--vanishing VEVs $\cal V$ to ensure a
supersymmetric vacuum.
Some of these fields will appear in the nonrenormalizable terms
(\ref{supterms}), leading to
effective operators of lower dimension. Their coefficients contain
factors of order ${\cal V}/M\sim1/10$.
Typically the solution of the D-- and F--flatness
constraints break most or all of the horizontal $U(1)$ symmetries. 

The question is whether there exist string symmetries,
which are beyond the GUT symmetries and can provide an appealing
explanation for the proton lifetime. In a beautifully insightful paper 
\cite{pati} Pati studied this question in the model of ref \cite{eu},
and showed that such symmetries indeed exist in the
string models. In ref. \cite{zprime} it was shown that the required
symmetries can in fact remain unbroken below the string scale, and
hence provide the needed suppression. However, as emphasized above, 
an important question is whether the additional symmetry that
suppresses the baryon and lepton number violating operators, 
still allows at the same time the generation of a neutrino 
seesaw mass matrix and hence an acceptable neutrino mass spectrum. 
In this paper we address this question. 

\section{Proton decay and superstring $Z^\prime$}

For concreteness we focus on the string model of ref. \cite{eu},
giving rise to the observable gauge symmetry
\beq
SU(3)\times SU(2)\times U(1)_C\times U(1)_{L}\times U(1)_{1}
\times\ldots\times U(1)_6.
\eeq
$U(1)_{C}$ and $U(1)_L$  will, apart from the SM's hypercharge, 
produce  one of the additional gauge bosons which appears in the 
string models, and constrain the proton decay mediating operators.
The charges of the Standard Model states under $U(1)_{L}$ and
$U(1)_C$ are fixed in the string model, as are the charges of
$U(1)_{1,...,6}$.
We further assume that the low energy spectrum contains
two light Higgs representations, which are the ${\bar h}_1$ and
$h_{45}$ of \cite{eu}, with
\begin{eqnarray}
H^{{{U}}}&=&\overline{h}_1~=~[(\overline{1},0);
(\overline{2},1)]_{0,-1,0,0,0,0},\nonumber\\
H^{{{D}}}&=&{h}_{45}~=~[(1,0);(2,-1)]_{-\frac{1}{2},-\frac{1}{2},0,0,0,0}.
\end{eqnarray} 
Prior to rotating the anomaly into a single $U(1)_{6'}$, all of the
$U(1)_{1,\cdots,6}$ are anomalous:
${\rm Tr} U_1= {\rm Tr} U_2={\rm Tr} U_3=24,~{\rm Tr} U_4= {\rm Tr} U_5=
{\rm Tr} U_6=-12$. These can be expressed by one anomalous
combination which is unique and five non--anomalous
ones:\footnote{The normalization of the
different $U(1)$ combinations is fixed
by the requirement that the conformal dimension of the
massless states gives ${\bar h}_1=1$ in the new basis. 
Here we neglect the normalization factors as we are only
interested in the invariant superpotential terms.}
\begin{eqnarray}\label{anomrot}
\left(\begin{array}{cccccc} 
U(1)_{1'} \\U(1)_{2'}\\U(1)_{3'}\\U(1)_{4^\prime}\\U(1)_{5'}\\U(1)_{6'} 
\end{array}\right)=
\left(\begin{array}{rrrrrr}
1  & -1 &  0 &  0  &  0  &  0\\
1  &  1 & -2 &  0  &  0  &  0\\
0  & 0  &  0 &  1  &  -1 &  0\\
0  & 0  &  0 &  1  &  1  & -2\\
1 & 1 & 1 & 2 & 2 & 2 \\
2 & 2 & 2 & -1& -1& -1
\end{array}\right) 
 \left(\begin{array}{cccccc} 
U(1)_1 \\U(1)_2\\U(1)_3\\U(1)_4\\U(1)_5\\U(1)_6 
\end{array}\right);\label{u16p}
\end{eqnarray}
The Standard Model weak hypercharge $U(1)_Y$ and the orthogonal
$U(1)_{Z^\prime}$ combinations are given by
\begin{eqnarray}
\left(\begin{array}{rr} 
U(1)_{Y} \\U(1)_{Z'}
\end{array}\right)=
\left(\begin{array}{cccccc}
\frac{1}{3}  & \frac{1}{2}\\
1&-1~~\\
\end{array}\right) 
\left(\begin{array}{cccccc} 
U(1)_{C} \\U(1)_L\end{array}\right).\label{u1yzp}
\end{eqnarray}
The complete massless spectrum and the charges under the four dimensional
gauge group are given in ref. \cite{eu}.
After the rotation (\ref{u1yzp}) one gets the charges

\beqn
\begin{tabular}{|c|rrrrrrrr|}
\hline
\bf{Field} & $\overline{E}$ & $\overline{U}$ & $Q$ & $\overline{N}$ &
$\overline{D}$ & $L$ & $H^U$ & $H^D$ \\
\hline
$U(1)_Y$ & $1$ & $-\frac{2}{3}$ & $\frac{1}{6}$ & $ 0
$ &$\frac{1}{3}$& $-\frac{1}{2}$& $ \frac{1}{2} $ & $-\frac{1}{2}$\\
\hline
$U(1)_{Z'}$ & $\frac{1}{2}$ & $\frac{1}{2}$ &$\frac{1}{2}$ & $\frac{5}{2}$ & 
$-\frac{3}{2}$ & $-\frac{3}{2}$ & $-1$ & $1$\\
\hline
\end{tabular}
\label{table1}
\eeqn
in agreement with the hypercharges of the Standard Model.
Furthermore (\ref{anomrot}) leads to the charges 
\beqn
\begin{tabular}{|c|rrrrrr|}
\hline
\bf{Field}  & $ U(1)_{1'}$ & $ U(1)_{2'}$  & $U(1)_{3'}$ &
$U(1)_{4^\prime}$ & $U(1)_{5'}$& $U(1)_{6'}$       \\
\hline 
$Q_1,~\overline{D}_1,~\overline{N}_1$ & $\frac{1}{2}~~~~$  &
$\frac{1}{2}~~~~$ & $-\frac{1}{2}~~~~$ &$-\frac{1}{2}~~~~$ &
$-\frac{1}{2}~~~~$ &$ \frac{3}{2}~~~~$   \\ 
\hline
$L_1,~\overline{E}_1,~\overline{U}_1$ &  $\frac{1}{2}~~~~$  &
$\frac{1}{2}~~~~$ & $\frac{1}{2}~~~~$ &$\frac{1}{2}~~~~$ &
$\frac{3}{2}~~~~$ &$ \frac{1}{2}~~~~$    \\ 
\hline
$Q_2,~\overline{D}_2,~\overline{N}_2$ &  $-\frac{1}{2}~~~~$ &
$\frac{1}{2}~~~~$ &$\frac{1}{2}~~~~$&  $-\frac{1}{2}~~~~$ &
$-\frac{1}{2}~~~~$ &$ \frac{3}{2}~~~~$  \\ 
\hline
$L_2,~\overline{E}_2,~\overline{U}_2$ &   $-\frac{1}{2}~~~~$ &
$\frac{1}{2}~~~~$ &$-\frac{1}{2}~~~~$&  $\frac{1}{2}~~~~$ &
$\frac{3}{2}~~~~$ &$ \frac{1}{2}~~~~$  \\ 
\hline
$Q_3,~\overline{D}_3,~\overline{N}_3$ &$0~~~~$& $-1~~~~$   &
$0~~~~ $ &$ 1~~~~$ &$-\frac{1}{2}~~~~$&$\frac{3}{2}~~~~$   \\ 
\hline
$L_3,~\overline{E}_3,~\overline{U}_3$ &$0~~~~$& $-1~~~~$   &
$0~~~~ $ &$- 1~~~~$ &$\frac{3}{2}~~~~$&$\frac{1}{2}~~~~$   \\ 
\hline
$H^{U}$ & $1~~~~$ & $-1~~~~$ &  $0~~~~$ &  $0~~~~$  &  $-1~~~~$ &
$-2~~~~$  \\ \hline
$H^{D}$ & $0~~~~$ & $-1~~~~$ &  $0~~~~$ &   $0~~~~$  &  $-1~~~~$ &
$-2~~~~$\\
\hline  
\end{tabular}.
\label{u1extras}
\eeqn
$~$\\
We note that $U(1)_{1'},...,U(1)_{4^\prime}$ are generation-dependent.

The superpotential of the MSSM+$N\!+\!{\not\!\!R_p}$ is given by
\begin{eqnarray}\label{superpot}
W~~~=~~~\varepsilon^{ab}~\delta^{xy}~G^{(U)}_{ij}~Q^i_{ax}~H^{U}_b~
\overline{U}^{j}_y~&~&~\nonumber\\
+~\varepsilon^{ab}~\delta^{xy}~G^{(D)}_{ij}~Q^i_{ax}~H^{D}_b~
\overline{D}_y^j~&~&~\nonumber\\
+~\varepsilon^{ab}~G^{(E)}_{ij}~L^i_{a}~H^{D}_b~
\overline{E}^{j}~&~&~+~\varepsilon^{ab}~G^{(N)}_{ij}~L^i_{a}~
H^{U}_b~\overline{N}^{j}\nonumber\\
+~\varepsilon^{ab}~\mu~ H_a^{D}~H_b^{U}  ~&~&~+~\Gamma_{ij}~
\overline{N}^{i}~\overline{N}^{j}\nonumber\\
 & & \nonumber\\
+~\frac{1}{2}~\varepsilon^{ab}~\Lambda_{ijk}~L^i_a~L^j_b~
\overline{E}^k~&~&\nonumber\\
+~\varepsilon^{ab}~\delta^{xy}~\Lambda^{'}_{ijk}~Q^i_{ax}~L^j_{b}~
\overline{D}^{k}_y~&~&~+~\Xi_i~\overline{N}^{i}   \nonumber\\
+~\frac{1}{2}~\varepsilon^{xyz}~\Lambda_{ijk}^{''}~\overline{U}^{i}_x~
\overline{D}^{j}_y~\overline{D}^{k}_z~&~&~+~\varepsilon^{ab}~{\Upsilon}_i~
\overline{N}^{i}~H^{D}_a ~H^{U}_b \nonumber\\
+~\varepsilon^{ab}~K_i~L_a^i~H^{U}_b~&~&~+~\Lambda^{'''}_{ijk}~
\overline{N}^{i}~\overline{N}^{j}~\overline{N}^{k}.\end{eqnarray}
Here the particle content is given by the MSSM--superfields
plus three generations of right--handed neutrinos. 
$Q,L,U,D,N,E$ represent the two left--handed $SU(2)$--doublets of the quarks
and leptons, and the  right--handed up--quarks, down--quarks,
neutrinos and electrons, respectively; $H^{D},H^{U}$ are the two
left--handed $SU(2)$ Higgs doublets; an overbar denotes
charge conjugation. The gauge group is $SU(3)\times SU(2) \times U(1)_Y$.
$a,b$ are $SU(2)$--indices (taking on the values $1,2$),
$~x,y,z$ are $SU(3)$--indices (taking on the values $1,2,3$),
$~i,j,k$  are generational indices, taking on the values $1,2,3$;
\emph{the heaviest generation is labeled by $1$, which is 
\underline{not} the same convention as e.g. in some papers 
on $\not\!\!R_p$; concerning $\Lambda_{ijk}'$, note that we
work with the operator $QL\overline{D}$ rather than $LQ\overline{D}$.}
Summation over repeated indices is implied;  $\delta^{...}$ is the
Kronecker symbol,
$\varepsilon^{...}$ symbolizes any  tensor that is totally antisymmetric
with respect to the exchange of two indices, with $\varepsilon^{12...}=1$.
All other symbols are coupling constants. The right blocks of (\ref{superpot})
contain right--handed neutrinos, the left ones do not; the upper blocks
contain
$R_p$--conserving terms, the lower ones do not. Thus the
superpotential of the MSSM is given by the upper left block. 

The unbroken new symmetries $U(1)_{1',\cdots,6'}$ strongly
reduce the number of
renormalizable operators in the low-energy superpotential
of the   MSSM+$\!N+\!{\not\!\!R_p}$:
Imposing
$U(1)_{Z'}\times U(1)_{...^\prime}$ on (\ref{superpot}) gives that
\begin{eqnarray}\label{theu}
U(1)_{Z'}\times U(1)_{1'}&~&\mbox{allows}~~G^{(U)}_{22},
G^{(N)}_{22},G^{(D)}_{12},G^{(D)}_{21},G^{(D)}_{33},G^{(E)}_{12},
G^{(E)}_{21},G^{(E)}_{33},\nonumber\\
U(1)_{Z'}\times U(1)_{2'}&~&\mbox{allows}~~G^{(U)}_{ij},G^{(N)}_{ij},
G^{(D)}_{ij},G^{(E)}_{ij}~\mbox{for}~i,j=1,2,\nonumber\\
U(1)_{Z'}\times U(1)_{3'}&~&\mbox{allows}~~G^{(U)}_{ii},
G^{(N)}_{ii}~\mbox{for}~i=1,2,3;\nonumber\\ 
                          &~& ~~~~~~~~~~G^{(D)}_{12},G^{(D)}_{21},
G^{(D)}_{33},G^{(E)}_{12},G^{(E)}_{21},G^{(E)}_{33},\mu,\nonumber\\
U(1)_{Z'}\times U(1)_{4^\prime}&~&\mbox{allows}~~G^{(U)}_{ii},
G^{(N)}_{ii}~\mbox{for}~i,j=1,2,3;\nonumber\\
                          &~&
{}~~~~~~~~~~G^{(U)}_{21},G^{(U)}_{12},
G^{(N)}_{21},G^{(N)}_{12};~\mu.\nonumber\\
U(1)_{Z'}\times U(1)_{5'}&~&\mbox{allows}~~G^{(U)}_{ij},
G^{(N)}_{ij}~\mbox{for}~i,j=1,2,3,\nonumber\\
U(1)_{Z'}\times U(1)_{6'}&~&\mbox{allows}~~G^{(U)}_{ij},
G^{(N)}_{ij}~\mbox{for}~i,j=1,2,3.
\end{eqnarray}
If  $U(1)_{Z^\prime}$ and \emph{all} $U(1)_{...'}$ were exact, only
the two terms $Q^2H^U\overline{U}^2$ and $L^2H^U\overline{N}^2$
in (\ref{superpot}) would be allowed. Thus we assume that all but
one, labeled $U(1)_{?^\prime}$, get broken near the heterotic--string
scale $M_{S}\sim10^{18}{\rm GeV}$.
Hence the gauge group at high energies below the Planck--scale is given by
\begin{eqnarray}\label{hegg}
SU(3)\times SU(2)\times U(1)_Y \times U(1)_{Z'} \times U(1)_{?^\prime}.
\end{eqnarray}
We assume that the $\overline{N}^i$ do not get a VEV. Thus to
break the $U(1)_{Z'}\times U(1)_{?'}$ gauge
symmetries we add the left--handed SM--singlet fields
$A,B$ and $R,S,T$ to the fields of the MSSM$+N$.
The underlying $SO(10)$ structure of the model
demands $A,B$ to be a vector--like couple with respect to $U(1)_{Z^\prime}$,
with $A$ having the charge of the right--handed neutrino $\overline{N}$.
We take   $A,B$  to be uncharged with respect to $U(1)_{?'}$, as we 
require that the VEVs of $A,B$ do not break $U(1)_{?'}$
because we assume the breaking scale of $U(1)_{Z'}$ to be much higher than
the breaking scale of $U(1)_{?'}$. 
The phenomenological constraints that we discuss below imply that
$\Lambda_{?^\prime}$ has to be considerably above the electroweak scale.
The stringy background of the model requires $R,S,T$ to be uncharged under
$U(1)_{Z'}$. Fixing their $U(1)_{?^\prime}$--charges basically is the only
freedom in our string--inspired model. The choices we make here aim at
generating a simple see--saw matrix.
In the string models the additional Higgs spectrum may arise
{}from fundamental states in the massless string spectrum,
or from product of fields that produce these effective
quantum numbers \cite{fh}.

{}From (\ref{theu}) we see that
$U(1)_{4^\prime},~U(1)_{5'},~U(1)_{6'}$ are too
restrictive, and $U(1)_{1'}$ still allows too few entries in $G^{(U)},G^{(N)}$
to be $U(1)_{?^\prime}$.  $U(1)_{2'}$ and $U(1)_{3'}$ look more promising,
and we study both cases.
$U(1)_{2'}$ leads to a model with no mass--scale, as it forbids the
$\mu$--term. If $U(1)_{2^\prime}$ suppresses, or forbids, the proton decay
mediating operators, while still allowing the see--saw mass terms, and setting

the  order of the $\mu$--parameter by its breaking scale, then 
$U(1)_{2^\prime}$ yields an example of a $U(1)$ symmetry
that, provided that it remains unbroken down to a sufficiently
low scale,  can address the proton lifetime and neutrino
mass issues, as well as provide a solution to the $\mu$--problem.
However, $U(1)_{2^\prime}$ also forbids the mass terms
for one generation, which are therefore suppressed by
$\Lambda_{2^\prime}/M_{S}$. One could contemplate the possibility that
this deficit is cured by radiative corrections. 
$U(1)_{3'}$ allows both masses for all generations and
also a $\mu$--term. In this case therefore we have to 
ascertain that the additional Higgs VEVs do not generate a
large $\mu$--term.  
However, unlike $U(1)_{2^\prime}$, $U(1)_{3^\prime}$
does not forbid the proton--destabilizing dimension five operators, 
and in particular, the operators $Q^3Q^3Q^3L^3$
and $U^3U^3D^3E^3$. Therefore in the
case of $U(1)_{3^\prime}$ additional suppression of the dimension
five operators is required. In the context of the string models
such suppression is anticipated due to the additional $U(1)$ symmetries
that are broken by the choices of supersymmetric flat directions
at the string scale.

\subsection{$\underline{\hbox{The case of $U(1)_{3^\prime}$}}$} 

We first focus on the case of $U(1)_{Z^\prime}\times U(1)_{3^\prime}$. 
We work with the following charges:\\
\beqn
\begin{tabular}{|c|rr|rrr|}
\hline
\bf{Field} & $A$ & $B$ & $R$ & $S$ & $T$ \\
\hline
$U(1)_{Z'}$ & $\frac{5}{2}$ & $-\frac{5}{2}$ & $0$ & $0$ & $~0$ \\
\hline
$U(1)_{3'}$ &  $0$ & $0$ & $\frac{1}{2}$ & $-\frac{1}{2}$ & $~0$\\
\hline
\end{tabular}
\label{extrafields}
\eeqn
As $T$ is completely uncharged
it cannot break any symmetry and we therefore impose $\langle T \rangle=0$. 
Instead of (\ref{superpot}) we now obtain
\begin{eqnarray} 
W&=&
\varepsilon^{ab}~\delta^{xy}~G^{(U)}_{ij}~Q^i_{ax}~H^{U}_b~
\overline{U}^{j}_y~+~
\varepsilon^{ab}~\delta^{xy}~G^{(D)}_{ij}~Q^i_{ax}~H^{D}_b~
\overline{D}_y^j~
\nonumber\\&+&~
\varepsilon^{ab}~G^{(E)}_{ij}~L^i_{a}~H^{D}_b~\overline{E}^{j}~+~
\varepsilon^{ab}~G^{(N)}_{ij}~L^i_{a}~H^{U}_b~\overline{N}^{j}~+~
\varepsilon^{ab}~\mu~ H^D_a~H^U_b
\nonumber\\~&+&~
\varepsilon^{ab}~P~T~H^D_a H^U_b~+~
\varepsilon^{ab}~\Upsilon^{'}~A~L^3_a~H_b^U~+~
\Theta~T~+~M_N~B~\overline{N}^3~
+~M_R~R~S\nonumber\\
&+&M_A~A~B~+~M_T~T~T~+~\Upsilon^{''}~B~R~\overline{N}^1
{}~+~\Upsilon^{'''}~B~S~\overline{N}^2~+~\Upsilon^{''''}~B~T~\overline{N}^3
\nonumber\\
&+&{P}'~A~B~T~+~P''~R~S~T~+~P'''~T~T~T,~~~~~~~~~
\label{renorsup}
\end{eqnarray}
with the $G^{(...)}_{ij}$ having only three entries  (see (\ref{theu})).
Guided by the structure of the string models we impose $\Theta=0$.
{}From the superpotential above we obtain
the following see--saw mass matrix ($n_L^3$ 
being the fermionic component of $L^3=N_L^3$,
$a$ being the fermionic component of $A$, etc.),  
\begin{eqnarray}\label{seesaw}
\left(\begin{array}{cccccccccccc}n_L^1~~~&n_L^2~~~&n_L^3~~~&
\overline{n_R}^1~~~&\overline{n_R}^2~~~&
\overline{n_R}^3~~~&r~~~&s~~~&t~~~&a~~~&b~~~\end{array}
\right)~~~~~~~~~~~~~~\nonumber\\
\left(\begin{array}{ccccccccccc} & & &\langle H^U\rangle& & & & & & & \\
           & & & &\langle H^U\rangle& & & & & & \\
           & & & & &\langle H^U\rangle & & & & \langle H^U \rangle& \\
   \langle H^U\rangle& & & & & &\langle B \rangle & & & &\langle R \rangle \\
   &\langle H^U \rangle& & & & & &\langle B \rangle & & &\langle S \rangle \\
   & & \langle H^U \rangle & & & & & &\langle B \rangle  & &M_N \\
   & & &\langle B \rangle &  & & &M_R &\langle S\rangle & & \\
   & & & &\langle B \rangle & & M_R & &\langle R\rangle & & \\
   & & & & &\langle B \rangle &\langle S\rangle &\langle R\rangle 
                 &M_T &\langle B \rangle &\langle A \rangle \\
   & &\langle H^U\rangle & & & & & &\langle B \rangle & &M_A \\
   & & &\langle R \rangle &\langle S 
    \rangle &M_N & & &\langle A \rangle 
                         &M_A &\end{array}\right)
\left(\begin{array}{c}n_L^1\\ n_L^2\\ n_L^3\\\overline{n_R}^1\\
\overline{n_R}^2\\\overline{n_R}^3\\r\\s\\t\\a\\b\end{array}\right),
\nonumber\\
\label{seesaw1}
\end{eqnarray}
where in (\ref{seesaw1}) coefficients of order one are not displayed 
explicitly, and the $n_L^3a$--entry is discussed further below.
The entries in  (\ref{seesaw1}) are constrained by several
phenomenological considerations. In order to avoid $D$--term
SUSY--breaking contributions that are larger than ${\rm 1~TeV }$
we impose
$
{}~\Big|\langle R \rangle^2-\langle S \rangle^2\Big|~\leq~ 1~({\rm TeV})^2~$
and
$~\Big|\langle A \rangle^2-\langle B \rangle^2\Big|~\leq~1~({\rm TeV})^2.
$ 
As the $U(1)_{Z^\prime}$ and $U(1)_{3^\prime}$ breaking scales are
assumed to be above the SUSY breaking scale we thus have
\beq\label{A=B}
\langle A \rangle \sim \langle B\rangle,~~\langle R
\rangle \sim \langle S \rangle.
\eeq

After $U(1)_{Z^\prime}$ breaking the term
$\Upsilon^{'}~A~L^3~H^{U}$ may generate a very large effective
$K_3$, see (\ref{superpot}). One might argue that it is possible to rotate
this term away by a unitary field-redefinition of $L^3$ and $H^D$, so that
\begin{eqnarray}
\left(\begin{array}{cc}\mu&K_3\end{array}\right)\left(\begin{array}
{c}H^D\\L^3\end{array}\right)
\longrightarrow~\left(\begin{array}{cc}\widetilde{\mu}&0
\end{array}\right)\left(\begin{array}{c}\widetilde{H}^D\\
\widetilde{L}^3\end{array}\right),
\end{eqnarray}
but this would induce a very large $\widetilde{\mu}$, namely
$|\widetilde{\mu}|^2=|\mu|^2+|\Upsilon^{'}~\langle A\rangle|^2$,
which is not desired, as $\mu\ll \langle A\rangle $. We therefore have to
demand that
\begin{equation}\label{314}
\Upsilon^{'}\leq\frac{\langle H^U \rangle }{ \langle A \rangle }.
\end{equation}
Thus, in the see--saw mass matrix ${\bf M}$ in (\ref{seesaw}) we can neglect
the
$n^3_La$-entries, as
${\langle H^U \rangle^2 }/{ \langle A \rangle }\ll\langle H^U \rangle$.

We impose the following relation on the parameters appearing
in (\ref{seesaw1}) which will yield appealing neutrino mass and mixing
spectrum
\begin{equation}\label{AlM}
M_N\sim \frac{\langle A \rangle^\frac{3}{2}}
{\langle R\rangle^\frac{1}{2}},
\label{equationabove}
\end{equation}
with $M_N$ much larger than $\langle A \rangle$.
Equation (\ref{equationabove}) is constrained by requiring no
$F$--term SUSY--breaking contributions that are bigger than
{1TeV}, i.e. $M_N\langle A \rangle \leq {\rm 1TeV} \times M_{S}$.
Similar constraints on products
like $\langle A \rangle\langle R \rangle$ are automatically fulfilled
with the condition guaranteeing no large contributions to the $\mu$--term,
that we discuss further below.   

We must also insure that the dangerous proton decay mediating
operators are sufficiently suppressed. Such operators
may arise from non--renormalizable terms of the form 
of\footnote{$k=1,2$, because
$k=3$ does not have to be considered due to the antisymmetry of
$\Lambda_{ijk}''$ under the exchange of $j,k$.}
\beq\label{op}
\frac{A~R}{M^2_{S}}~ Q^3~L^3~\overline{D}^k~~\mbox{and}~~\frac{A~R}
{M^2_{S}}~ \overline{U}^3~\overline{D}^3~\overline{D}^k.
\eeq
After $U(1)_{Z'}$  and $U(1)_{3'}$ symmetry breaking one  gets
\beq\label{llar}
\Lambda_{33k}'~\mbox{and}~\Lambda_{33k}''\sim \frac{\langle A \rangle~\langle
R
\rangle}{M_{S}^2}.
\eeq
Proton decay limits impose that (see for example \cite{herbi})
\begin{eqnarray}\label{proyon}
\Lambda^{'}_{33k}~\Lambda^{''}_{33k}~\leq~2\times10^{-27}\Bigg(
\frac{m_{\widetilde{d_R^k}}}{100~{\rm GeV}}\Bigg)^2;\end{eqnarray}  
taking the masses of the SUSY-particles to be of $O(1{\rm TeV})$,
one gets that the relation above is fulfilled if
\begin{eqnarray}
\Lambda^{'}_{33k}\leq 5\times 10^{-13}~~\mbox{and}~~
\Lambda^{''}_{33k}\leq 5\times 10^{-13}.
\end{eqnarray}
Comparing this with eq. (\ref{llar}) gives the condition
\beq
\langle A \rangle ~\langle R \rangle~ \leq~ 5\times 10^{23} ({\rm GeV})^2.
\eeq
 
Similarly, potential terms contributing to the electroweak Higgs
mixing parameter should be adequately suppressed or absent altogether. 
As the VEV of $T$ vanishes, the superpotential term
$TH^DH^U$ does not generate a large $\mu$--term.
To guarantee that the two gauge--invariant non--renormalizable terms
\begin{equation}
\frac{A~B}{M_{S}}~H^D~H^U,~~\frac{R~S}{M_{S}}~H^D~H^U
\end{equation}
do not contribute substantially to the $\mu$--term after $U(1)_{Z^\prime}$
and $U(1)_{3^\prime}$ gauge symmetry breaking, we impose 
\begin{equation}\label{Acon}
\langle A \rangle\leq 10^{10} {\rm GeV}.
\label{threetwothree}
\end{equation} 
Since we have $\langle A \rangle\gg\langle R \rangle$
eq. (\ref{threetwothree}) implies
\begin{eqnarray}
\langle A \rangle\langle R \rangle\leq 10^{20} ({\rm  GeV})^2 ,\end{eqnarray}
so that  the operators (\ref{op}) are rendered harmless, and also $F$--term
SUSY--breaking is avoided.

As stated earlier, the $U(1)_{2'}$/$U(1)_{3^\prime}$--charges of the
three generations are not family universal.
Consequently, the breaking
scale of $U(1)_{2'}$/$U(1)_{3^\prime}$ has to be sufficiently high to avoid
contradiction
with Flavor Changing Neutral Currents (FCNC):
\begin{eqnarray}\label{fcnc}
\langle R \rangle \geq 30\times10^{3}{\rm GeV}.
\end{eqnarray}

For simplicity we assume\footnote{Other choices
either did not lead to substantially different results or produced
contradiction with experiment.}
\beq\label{8}
M_R\sim M_T\sim \langle R \rangle,~~M_A\sim \langle A \rangle.
\eeq

We plug (\ref{A=B}), (\ref{314}), (\ref{AlM}) and (\ref{8})
into (\ref{seesaw}) and  determine its eigenvalues. The eight large ones are
approximately given by
\begin{equation}
\pm M_N,\pm\langle A\rangle,\pm\langle A\rangle,\pm\langle A\rangle,
\end{equation}
all large enough that the corresponding particles cannot have been observed.
However, we are of course most interested in the three small
eigenvalues. 
Writing
\begin{eqnarray}
{\bf H}=\left(\begin{array}{cccccccc}
\langle H^U \rangle & 0 & 0 & 0 & 0 & 0 & 0 & 0 \\
0 & \langle H^U \rangle & 0 & 0 & 0 & 0 & 0 & 0 \\
0 & 0 & \langle H^U \rangle & 0 & 0 & 0 & 0 & 0\end{array}\right)
\end{eqnarray}
and 
\begin{eqnarray}
{\bf J}=\left(\begin{array}{cccccccc} 
  & & &\langle A \rangle & & & &\langle R \rangle \\
  & & & &\langle A \rangle & & &\langle R \rangle \\
  & & & & &\langle A \rangle  & &\frac{\langle A \rangle^\frac{3}{2}}
  {\langle R \rangle^\frac{1}{2}} \\
  \langle A \rangle & & & &\langle R \rangle &\langle R \rangle & & \\
  &\langle A \rangle& &\langle R \rangle & &\langle R \rangle & & \\
  & &\langle A \rangle &\langle R \rangle &\langle R \rangle &\langle R 
  \rangle &\langle A \rangle &\langle A \rangle \\
  & & & & & \langle A \rangle& &\langle A\rangle \\
  \langle R \rangle  &\langle R \rangle &\frac{\langle A
\rangle^{\frac{3}{2}}}
  {\langle R \rangle^\frac{1}{2}} & & &\langle A \rangle &
  \langle A\rangle & \end{array}\right),
\end{eqnarray}
we can express the see--saw mass matrix as
\begin{equation}\label{svss}
{\bf M}=\left(\begin{array}{cc}0&{\bf H}\\{\bf H}^T&{\bf J}\end{array}\right).
\end{equation}
Thus the three lightest eigenvalues and their corresponding eigenvectors
are approximately, and to lowest order in 
${\langle R \rangle}/{\langle A \rangle}$, those of the $3\times3$ matrix
\begin{equation}
{\bf{H}}^T{\bf{J}}^{-1}{\bf{H}}~\approx~\frac{\langle H^U\rangle^2
\langle R \rangle}{\langle A\rangle^2}~\left(\begin{array}{ccc}0&1
&0\\
1 & 0&0\\
0&0&-2 
\end{array}\right),
\label{lightneutrinomassmatrix}
\end{equation}
namely \footnote{This result can be found by an alternative 
method which for simplicity we demonstrate only for the simple conventional
$2\times2$ see--saw matrix, which gives the characteristic
polynomial $x^2-Mx-m^2=0$. Assuming that $|x^2|\ll |Mx|,m^2$ one has
$~~-Mx-m^2\approx 0~~\mbox{so that}~~x\approx-\frac{m^2}{M}$,
reproducing the well-known result. Plugging it into the initial
assumption above gives $\frac{m^4}{M^2}\ll M\frac{m^2}{M},m^2$,
justifying the assumption in hindsight.} 
\begin{equation}
\pm~\frac{\langle H^U \rangle^2\langle R \rangle}{\langle A
\rangle^2},~~\frac{2\langle H^U \rangle^2\langle R \rangle}{\langle A
\rangle^2}.
\label{eigenvalues}
\end{equation}
The mass--degeneracy of the first two eigenvalues can be lifted by including 
higher orders, and/or non--universal coefficients, to obtain (\ref{deltam}).
All expressions go to zero for vanishing Higgs VEV, as they should.
The diagonalizing matrix of (\ref{lightneutrinomassmatrix}) is given by 
\begin{equation}
\left(\begin{array}{ccc}\frac{1}{\sqrt{2}} &-\frac{1}{\sqrt{2}}&0\\ 
\frac{1}{\sqrt{2}} & \frac{1}{\sqrt{2}}&0\\ 0 & 0 & 1\end{array}\right)
\left(\begin{array}{c} n_L^1\\ n_L^2\\ n_L^3\end{array}\right)=
\left(\begin{array}{c}\nu_1\\\nu_2\\\nu_3 
\end{array}\right).
\end{equation}
Here $\nu$ symbolizes the light mass--eigenstates. This result can be
interpreted as the (currently disfavored by SNO data \cite{sno})
maximal atmospheric neutrino mixing \cite{superk}
($\vartheta_{23}\sim45^\circ$), and
small mixing angle MSW solar neutrino oscillation
($\vartheta_{13}\sim\vartheta_{12}\sim 0^\circ$), i.e. one has
to fix the values of $\langle R \rangle$ and $\langle A \rangle$ with
\begin{eqnarray}\label{neumass}
\Big|m_2^2-m_3^2\Big|&=&3.5\times 10^{-3} ({\rm eV})^2,\\
\Big|m_3^2-m_1^2\Big|&=&3.5\times 10^{-3} ({\rm eV})^2,\\
\label{deltam}\Big|m_1^2-m_2^2\Big|&=&6\times 10^{-6} ({\rm eV})^2.
\end{eqnarray}
{}From the eigenvalues (\ref{eigenvalues})
we know that $m_3\sim 2m_2,2m_1$. Therefore, 
\beq
\frac{\sqrt{3}~\langle H^U\rangle^2~\langle R \rangle}
{\langle A \rangle^2}\sim\sqrt{3.5\times 10^{-3}({\rm{eV}})^2},
\eeq
and thus
\beq
\langle R \rangle \sim 3\times 10^{-15}({\rm GeV})^{-1} \langle A \rangle^2.
\eeq
{}From (\ref{fcnc}) we hence obtain $\langle 
A\rangle^2~\times~\frac{3~\times10^{-15}}{{\rm GeV}}~
\ge30~\times10^{3}{\rm GeV}$, so that together with
(\ref{Acon}) we have the narrow range
\begin{equation}
3\times 10^9{\rm GeV}\leq\langle A \rangle\leq 10^{10}{\rm GeV}.
\label{avev}
\end{equation}
Hence
\beq
30~\mbox{TeV}~\leq~\langle R \rangle \leq~300~\mbox{TeV},
\eeq
which leads to 
\beq
M_N \sim 10^{12}{\rm GeV},
\eeq
hence fulfilling the constraints of $F$--term SUSY
breaking with $\langle A\rangle=3\times10^9$GeV.

When $A$ acquires a VEV several renormalizable $\not\!\!R_p$ couplings
are generated from the terms 
\begin{equation}\label{rpvgen}
\frac{A}{M_{S}}~L~L~\overline{E},~~\frac{A}{M_{S}}~Q~L~\overline{D},~~
\frac{A}{M_{S}}~\overline{U}~\overline{D}~\overline{D}.
\label{lleqldudd}
\end{equation}
Assuming their coupling constants to be $O(1)$ one gets  
\begin{eqnarray}
&&\Lambda_{231},\Lambda_{132},\Lambda_{123},
\nonumber\\
&&\Lambda'_{231},\Lambda'_{132},\Lambda'_{113},\nonumber\\
&&\Lambda'_{311},\Lambda'_{322},\Lambda'_{223},\Lambda'_{333}\nonumber\\
&&\Lambda''_{113},\Lambda''_{223},\Lambda''_{312},\label{coupcon}
\end{eqnarray}
all being suppressed by a factor of
$\frac{\langle A\rangle}{M_{S}}\sim10^{-8}$,
which is below the current upper bounds at high
energies.\footnote{See Table III in \cite{herbi2}.
Note that their convention
is to label the heaviest generation `$3$',
and that they work with $LQ\overline{D}$ rather than
$QL\overline{D}$.} We also note that the lepton number violating
operators in (\ref{lleqldudd}) can contribute to the left--handed
neutrino masses through fermion--sfermion loops, with 
$m_\nu\sim3\Lambda^{\prime^2}/(16\pi^2)m_b^2/M_{\rm SUSY}$ \cite{sacha}.
However, with the constraint (\ref{avev}), this contribution is at
most $m_\nu\sim5\cdot10^{-11}{\rm eV}$ and therefore negligible.\\

\subsection{$\underline{\hbox{The case of $U(1)_{2^\prime}$}}$} 

Turning to $U(1)_{Z'}\times U(1)_{2'}$ we work with the following charge
assignment:\footnote{Being far from obvious we have not found any other
$U(1)_{2'}$--charge assignment for $R,S,T$ that produced a stable proton,
a sensible see--saw matrix and the generation of a $\mu$--term. Just as for
$U(1)_{3'}$, the $U(1)_{2'}$--charges of $R,S,T$ are opposite to the ones
of the neutrinos.}
\\
\beqn
\begin{tabular}{|c|rr|rrr|}
\hline
\bf{Field} & $A$ & $B$ & $R$ & $S$ & $T$ \\
\hline
$U(1)_{Z'}$ & $\frac{5}{2}$ & $-\frac{5}{2}$ & $0$ & $0$ & $~0$ \\
\hline
$U(1)_{2'}$ &  $0$ & $0$ & $-\frac{1}{2}$ & $-\frac{1}{2}$ & $~1$\\
\hline
\end{tabular}.
\label{extrafields2}
\eeqn
It allows the non--renormalizable term 
\begin{equation}\label{RR}
\frac{T~T~H^U~H^D}{M_{S}},
\end{equation}
so that we can generate the $\mu$--term with
\begin{equation}
\langle T \rangle^2\sim10^2~\mbox{GeV}\times M_{S}~~\Rightarrow~~
\langle T \rangle
\sim 10^{10}~\mbox{GeV}.
\end{equation}
Thus if $\mathcal{N},\mathcal{N}',\mathcal{N}''$ (see (\ref{spau}))
are $\leq 10$ the superpotential terms $RST,RRT,SST$ cannot
contribute substantially to $F$--term SUSY--breaking (see (\ref{Laeibl}) and
(\ref{fterm})).
Since the $U(1)_{2'}$--charge of ($H^UH^D$) is $-2$, and  neither a
three--field combination of $R,S,T$ or one of these fields alone has a
compensating $U(1)_{2'}$--charge of $+2$, the $U(1)_{2'}$--charges forbid 
\begin{eqnarray}\label{RRR}
\frac{(R~or~S~or~T)~(R~or~S~or~T)~(R~or~S~or~T)~H^U~H^D}{M_{S}^2}\nonumber\\
\mbox{and}~~\frac{A~B~(R~or~S~or~T)~H^U~H^D}{M_{S}^2}.~~~~~~~~~~~~
\end{eqnarray}
Furthermore, since
\begin{equation}\label{prot}
Q_R,Q_S,Q_T\neq\frac{3}{2},
\end{equation}
the terms
\begin{equation}
\frac{A~(R~or~S~or~T)~Q^3~L^3~\overline{D}^{1,2}}{M_{S}^2}~~
\mbox{and}~~\frac{A~(R~or~S~or~T)~\overline{U}^3~\overline
{D}^3~\overline{D}^{1,2}}{M_{S}^2},
\end{equation}
are forbidden (unlike the case with $U(1)_{3'}$) and thus the
proton is protected.
The dimension five operator $\frac{1}{M_{S}}Q^3Q^3Q^3L^3$
is also forbidden (again unlike $U(1)_{3^\prime}$).
It can be generated from
$\frac{T~T~T~T}{M_{S}^5}Q^3Q^3Q^3L^3$, which however is highly suppressed.
Similarly, all other dimension five operators are suppressed by at least
$\langle T\rangle/M_{S}$ and are therefore adequately suppressed.
The renormalizable superpotential is given by
\begin{eqnarray} \label{spau}
W~&=&~\varepsilon^{ab}~\delta^{xy}~G^{(U)}_{ij}~Q^i_{ax}~
H^{\mathcal{U}}_b~\overline{U}^{j}_y~+~\varepsilon^{ab}~
\delta^{xy}~G^{(D)}_{ij}~Q^i_{ax}~H^{\mathcal{D}}_b~\overline{D}_y^j~+~
\varepsilon^{ab}~G^{(E)}_{ij}~L^i_{a}~H^{\mathcal{D}}_b~
\overline{E}^{j}~\nonumber\\
&~&~+~\varepsilon^{ab}~G^{(N)}_{ij}~L^i_{a}~H^{\mathcal{U}}_b~
\overline{N}^{j}~+~M_A~A~B~+~\Upsilon^{'}_i~B~R~\overline{N}^i~+~
\Upsilon^{''}_i~B~S~\overline{N}^i\nonumber\\
&~&~+~\Upsilon^{'''}~B~T~\overline{N}^3~+~{\mathcal{N}}~R~S~T~+~
{\mathcal{N}}'~R~R~T~+~{\mathcal{N}}''~S~S~T,~~~~~~~~~~~~~~~~~~\end{eqnarray}
with $i,j=1,2$. Looking at the charges we see that we can avoid
$D$--term SUSY--breaking with
\begin{equation}\label{Laeibl}
\langle R\rangle=\langle S\rangle=\langle T\rangle,~~
\langle A\rangle=\langle B\rangle.
\end{equation}
Thus (\ref{fcnc}) is fulfilled.

As in the case of $U(1)_{3'}$ we have $\langle R\rangle\ll\langle A\rangle.$
$M_A$ and $\langle A \rangle$ have to be chosen such that
the SUSY--breaking $F$--terms are adequately suppressed.
Thus, we impose
\begin{eqnarray}\label{fterm}
M_A ~\langle A \rangle &\leq& 1~\mbox{TeV}\times M_{S};\nonumber\\
\Big(\Upsilon'_i~\mbox{and}~\Upsilon''_i~\mbox{and}~\Upsilon'''\Big)~
\langle R \rangle ~\langle A \rangle &\leq &1~\mbox{TeV}\times M_{S}
\nonumber\\
\Rightarrow~\langle A \rangle &\leq & \frac{10^{11}~\mbox{GeV}}{\Big
(\Upsilon'_i~\mbox{and}~\Upsilon''_i~\mbox{and}~\Upsilon'''\Big)}.
\end{eqnarray}

The see--saw matrix is again of the form (\ref{svss}), with  
\begin{eqnarray}
{\mathbf{H}}=\left(\begin{array}{cccccccc}
\langle H^U \rangle & \langle H^U \rangle & 0 & 0 & 0 & 0 & 0 & 0 \\
\langle H^U \rangle & \langle H^U \rangle & 0 & 0 & 0 & 0 & 0 & 0 \\
0 & 0 & 0 & 0 & 0 & 0 & 0 & 0\end{array}\right)
\end{eqnarray}
and 
\begin{eqnarray}
{\mathbf{J}}=\left(\begin{array}{cccccccc} 
                                & & &\langle A \rangle &\langle A
\rangle & & &\langle R \rangle \\
                                & & &\langle A \rangle &\langle A
\rangle & & &\langle R \rangle \\
                                & & & & &\langle A \rangle  & &
{\langle R \rangle} \\
                                \langle A \rangle &\langle A
\rangle & &\langle R \rangle &\langle R \rangle &\langle R \rangle & & \\
                              \langle A \rangle  &\langle A
\rangle& &\langle R \rangle &\langle R \rangle &\langle R \rangle & & \\
                                & &\langle A \rangle &\langle R
\rangle &\langle R \rangle & & & \\
                                & & & & & & &M_A \\
                           \langle R \rangle     &\langle R \rangle & 
{\langle R \rangle}  & & & &M_A &\end{array}\right),
\end{eqnarray}
where we have not displayed explicitly 
the coupling constants $G^{(N)}_{ij}$,
 $\Upsilon'_i$, $\Upsilon''_i$, $\Upsilon'''$, $\mathcal{N}$,$\mathcal{N}'$,
 $\mathcal{N}''$ \footnote{These couplings
however cannot be completely random,
otherwise the matrix could be singular, e.g. all couplings exactly equal
to unity does not work, unlike the case for $U(1)_{3'}$.}.
The eight large  eigenvalues
corresponding to heavy particles are approximately
\begin{equation}
 \pm\langle A\rangle, \pm\langle A\rangle, \pm\langle A\rangle,\pm M_A.
\end{equation}
It turns out that to a good approximation the small eigenvalues of the 
see--saw matrix, corresponding to the very light neutrinos, are independent of
$M_A$, one has
\begin{equation}
{\mathbf{H}}^T{\mathbf{J}}^{-1}{\mathbf{H}}~\approx~\frac{\langle
H^{U}\rangle^2\langle R \rangle}{\langle A\rangle^2}~
\left(\begin{array}{ccc}\alpha&\beta &0\\
\beta & \gamma&0\\
0&0&0 \end{array}\right),
\end{equation}
$\alpha,\beta,\gamma$ being complicated functions of $G^{(N)}_{ij},
\Upsilon'_i, \Upsilon''_i, \Upsilon''', \mathcal{N},\mathcal{N}',
\mathcal{N}''$. 
{}From the null--row in the matrix above it follows that one of the neutrinos
is massless. It is possible to assign values between $\pm 0.1$ and
$\pm 10$ to  $G^{(N)}_{ij}, \Upsilon'_i, \Upsilon''_i, \Upsilon''',
\mathcal{N},\mathcal{N}',\mathcal{N}''$ such that one gets nearly
degenerate masses for the other two neutrinos, and maximal mixing
between them, obeying (\ref{bed}). In order for this to reproduce
the experimental data (\ref{neumass}) with $m_3\sim0$ one thus needs
\begin{equation}
m_{1,2}=
\pm\frac{\langle H^U\rangle^2\langle R \rangle}{\langle A\rangle^2}=
\pm 6\times10^{-11}\mbox{GeV}~~\Rightarrow~~\langle A\rangle\sim
10^{12}\mbox{GeV}.
\end{equation} 
In order to avoid substantial contributions to
SUSY--breaking $F$--terms we thus have to impose (see (\ref{fterm}))
\begin{equation}\label{bed}
\Upsilon'_i,\Upsilon''_i,\Upsilon'''~\leq 10^{-1},
\end{equation}
and $M_A\leq10^{9}$GeV\footnote{However, $M_A$ should be
sufficiently large so that the two particles
with mass $\sim M_A$ would not have been detected}.
{}From the terms (\ref{rpvgen}) one gets
\begin{eqnarray}
&&\Lambda_{131},\Lambda_{231},\Lambda_{132},\Lambda_{232},\Lambda_{123};
\nonumber\\
&&\Lambda'_{131},\Lambda'_{231},\Lambda'_{132},\Lambda'_{232},
\Lambda'_{123},\Lambda'_{113},\nonumber\\
&&\Lambda'_{311},\Lambda'_{321},\Lambda'_{312},\Lambda'_{322},
\Lambda'_{213},\Lambda'_{223};\nonumber\\
&&\Lambda''_{131},\Lambda''_{123},\Lambda''_{213},\Lambda''_{223},
\Lambda''_{312},
\end{eqnarray}
this time all of them are suppressed by
$\frac{\langle A \rangle}{M_{S}}=O(10^{-6})$,
again not contradicting the bounds of Table III in
\cite{herbi2}\footnote{We note again that in our convention `$1$' labels
the heaviest generation; we also work with $QL\overline{D}$ instead of
$LQ\overline{D}.$}. Similarly, we note that the fermion--sfermion
contribution to the left--handed neutrino masses \cite{sacha}
are adequately suppressed. 

\section{Discussion and Conclusions}

The structure of the Standard Model spectrum indicates the
realization of grand unification structures in nature. 
On the other hand the proton longevity severely constrains
the possible extensions of the Standard Model and
serves as a useful guide in attempts to understand the
origin of the Standard Model gauge and matter spectrum.
The realistic free fermionic heterotic--string models
reproduce the grand unification structures that are suggested
by the Standard Model and represent the most realistic
string models constructed to date. As such the realistic 
free fermionic models serve as a useful probe to the
fundamental characteristics of the possibly true 
string vacuum, as well as to various properties that
the string vacuum should possess in order to satisfy various
phenomenological constraints. It was proposed previously
that proton stability necessitates the existence of 
an additional $U(1)$ symmetry, which remains unbroken down to
intermediate or low energies. Furthermore, the required symmetry
is not of the type that arises in Grand Unified Theories,
but is of intrinsic string origin. The realistic free fermionic 
models do indeed give rise to $U(1)$ symmetries, which are
external to the GUT symmetries, and forbid the proton decay 
mediating operators. 

One of the vital issues in heterotic--string Grand Unification
is whether it is possible to adequately suppress the proton decay mediating
operators, and at the same time insure that the 
left--handed neutrino masses are sufficiently small.
That this is a non--trivial task is seen, for example,
from the fact that a VEV of the 
neutral component in the $16$ and $\overline{16}$
of $SO(10)$, which is used in the string models
to generate the required large see--saw mass scale, at the same
time can induce the dangerous proton decay mediating operators
from non--renormalizable terms. The solution pursued here therefore advocates
the existence of an additional $U(1)$ symmetry, which forbids the
proton decay mediating operators, while it permits the 
neutrino seesaw--matrix mass terms, as well as
those of the quarks and charged--leptons. Thus, provided
that the additional $U(1)$ is broken at a sufficiently low scale,
the proton decay mediating operators will be adequately suppressed.
In this paper we demonstrated that indeed the symmetries that
appear in the string models can guarantee proton longevity
and simultaneously suppress the left--handed neutrino masses.
We achieved this by assuming additional Higgs spectrum
that is needed to break the $U(1)_{Z^\prime}\times U(1)_{?^\prime}$
and by detailed construction of the neutrino see--saw mass matrix.
Such additional Higgs spectrum arises in the string models
from fundamental states or from product of states that produce
the effective quantum number of the states that we assumed here.
We then analyzed the neutrino see--saw mass matrix and
showed that the resulting neutrino mass spectrum is
in qualitative agreement with the experimentally observed values.

It is interesting to note that among the semi--realistic orbifold models
constructed to date the free fermionic models are unique in the 
sense that they are the only ones that have been shown to admit the $SO(10)$
embedding of the Standard Model spectrum. This is true of 
the $Z_3$ heterotic string models \cite{z3} as well as the type I
string models, which do not admit the chiral 16 of $SO(10)$
in the perturbative massless spectrum. This in particular means
that these models do not admit the canonical
$SO(10)$ embedding of the weak--hypercharge. In these
models the weak--hypercharge arises from a non--minimal
combination of the world--sheet $U(1)$ currents, which 
correctly reproduces the Standard Model hypercharge 
assignments. Therefore, the models do contain a number of 
$U(1)$ symmetries that may suppress the proton decay
mediating operators, and the suppression depends
on the $U(1)_{Z^\prime}$ symmetry breaking scale.
However, the non--canonical embedding of the weak
hypercharge typically results in exotic states which 
cannot be decoupled from the massless spectrum. As regard
the issue of neutrino masses, the non--standard embedding 
of the weak--hypercharge implies that this issue can only
be studied on a case by case basis, and general patterns 
are more difficult to decipher. The case of type I
constructions presents additional novelties. In this case
the Standard Model gauge group typically arises from a 
product of $U(n)$ groups. Therefore, embedding of the Standard Model
gauge group in an enlarged gauge structure
is not obtained in type I string models. Consequently,
one then often finds a gauged $U(1)_B$ in these models,
which guarantees proton longevity. The suppression of neutrino
masses in this context still remains an open issue that can
only be addressed on a case by case basis. However, due
to the non--standard embedding of the weak--hypercharge,
these models have typically been considered in the context
of low scale gravity models \cite{ibanezI}. In which case the suppression 
of of left--handed neutrino masses may arise from the large volume
of the extra dimensions \cite{lednm}.

To conclude, the Standard Model multiplet structure, augmented
with the right--handed neutrinos, strongly indicates the realization
of an underlying $SO(10)$ structure in nature. Preservations
of the $SO(10)$ embedding in the context of string unification
has proven to be a highly non--trivial task. In fact, among the 
semi--realistic orbifold models, only the free fermionic models
are known to reproduce this desired structure. However, the problems
of proton longevity and the simultaneous suppression of the left--handed
neutrino masses, still pose an extremely severe challenge to
Grand Unified, and heterotic--string, models. It has been proposed
that these challenges necessitate the existence of an extra non--GUT
$U(1)$ symmetry, broken at low energies, and that the required
symmetries do arise in the free fermionic heterotic--string models.
In this paper by constructing detailed string inspired models,
we showed that the additional stringy $U(1)$ symmetries can indeed perform
their designated task. The relative simplicity of the additional
Higgs spectrum that we assumed here, suggests that further more detailed
models might be able to fully reproduce the observed quark and lepton
mass spectrum, as well as safeguarding the proton lifetime.

\bigskip
\medskip
\leftline{\large\bf Acknowledgments}
\medskip
We would like to thank Sacha Davidson and Athanasios
Dedes for useful discussions
and the CERN theory group for hospitality.
This work was supported in part by PPARC. M.T. would like to thank
Bonn University for hospitality and
the Evangelisches Studienwerk and Worcester College for financial support.

\vfill\eject

\bibliographystyle{unsrt}

\end{document}